\documentclass[prl,aps,groupedaddress,twocolumn,floatfix]{revtex4}
\usepackage{amsmath,amsthm,amsfonts,amssymb,graphics,graphicx,epsfig,color,times,natbib,subfigure}

\newcommand{\ket}[1]{\mbox{$ | #1 \rangle $}}

\newcommand{\be}{\begin{equation}}
\newcommand{\ee}{\end{equation}}
\newcommand{\ba}{\begin{eqnarray}}
\newcommand{\ea}{\end{eqnarray}}

\newcommand{\one}{\leavevmode\hbox{\small1\normalsize\kern-.33em1}}

\begin{document}

\title{Quantifying multipartite nonlocality}
\author{Jean-Daniel Bancal, Cyril Branciard, Nicolas Gisin, Stefano Pironio}
\address{Group of Applied Physics, University of Geneva, 20 rue de l'Ecole-de-M\'edecine, CH-1211 Geneva 4, Switzerland}
\date{August 27, 2009}
%\maketitle

\begin{abstract}
The nonlocal correlations of multipartite entangled states can be reproduced by a classical model if sufficiently many parties join together or if sufficiently many parties broadcast their measurement inputs. The maximal number $m$ of groups and the minimal number $k$ of broadcasting parties that allow for the reproduction of a given set of correlations quantify their multipartite nonlocal content. We show how upper-bounds on $m$ and lower-bounds on $k$ can be computed from the violation of the Mermin-Svetlichny inequalities. While $n$-partite GHZ states violate these inequalities maximally, we find that W states violate them only by a very small amount.
\end{abstract}

\maketitle
By performing local measurements on an $n$-partite entangled state one obtains outcomes that may be nonlocal, in the sense that they violate a Bell inequality~\cite{bell}. Since the seminal work of Bell, nonlocality has been a central subject of study in the foundations of quantum theory and has been supported by many experiments \cite{genovese,aspect}. More recently, it has also been realized that it plays a key role in various quantum information applications \cite{barrett,brukner}, where it represents a resource different from entanglement. For instance, the security of device-independent quantum key distribution requires the existence of nonlocal correlations between the honest parties, very much in the spirit of Ekert's protocol~\cite{ekert,bhk,abgmps,masanes1}, and the only entanglement witnesses that do not rely on assumptions on the dimension of the Hilbert spaces are Bell inequalities, i.e., witnesses of nonlocality \cite{abgmps}.

While nonlocality has been extensively studied in the bipartite ($n=2$) and to a lesser extent in the tripartite ($n=3$) case, the general $n$-partite case remains much unexplored. The physics of many-particle systems, however, is well known to differ fundamentally from the one of a few particles and to give rise to new interesting phenomena, such as phase transitions or quantum computing. Entanglement theory, in particular, appears to have a much more complex and richer structure in the $n$-partite case than it has in the bipartite setting \cite{vidal,horod}. This is reflected by the fact that multipartite entanglement is a very active field of research that has led to important insights into our understanding of many-particle physics (see, e.g., \cite{fazzio,vidal2}). In view of this, it seems worthy to investigate also how nonlocality manifests itself in a multipartite scenario. What new features emerge in this context and what are their fundamental implications? How to characterize the nonlocality of experimentally realizable multi-qubit states, such as W states for instance? What role do $n$-partite nonlocal correlations play in quantum information protocols, e.g., in measurement-based computation \cite{mbqc}?

The vision behind the present paper is that in order to answer such questions and make further progress on our understanding of multipartite nonlocality, one should first find ways to \emph{quantify} it. Motivated by this idea, we introduce two simple measures that quantify the multipartite extent of nonlocality.

A natural way to characterize nonlocality is to attempt to replicate it using models where some non-local interactions (such as communication) are allowed between some parties. The first measure that we consider is based on classical communication models \`a la Svetlichny \cite{svetlichny,collins,NL,jones}, where the $n$ parties are divided into $m$ disjoint subgroups. Within each group the parties are free to collaborate and communicate with each other, but are not allowed to do so between distinct groups. The idea is that a given set of correlations contains more multipartite nonlocality if more parties need to join to be able to reproduce these correlations (see Fig.~1). The second measure of multipartite nonlocality that we introduce is based on models where $k$ parties broadcast their measurement inputs to all others. The idea again is that correlations that require more broadcasting parties to be simulated contain more multipartite nonlocality. The maximal number $m$ of groups and the minimal number $k$ of broadcasting parties that allow for the reproduction of a given set of correlations thus represent two simple ways of quantifying their multipartite nonlocal content.

\begin{figure}[h]
\begin{center}
\includegraphics[width=0.47\textwidth]{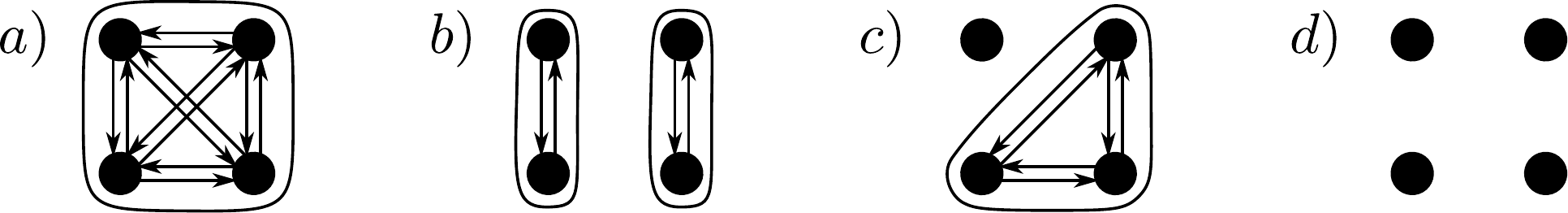}
\end{center}
\caption{Different groupings of $n=4$ parties into $m$ groups. Within each group, every party can communicate to any other party, as indicated by the arrows. $a)$ If all parties join into one group ($m=1$), they can achieve any correlations. $b)$ and $c)$ If they split into $m=2$ groups they can realize some non-local correlations but not all. $d)$ If they are all separated ($m=n$), they can only reproduce local correlations. }
\label{fig:4parties}
\end{figure}

Given an arbitrary set of correlations, it may in general be difficult to determine the corresponding values of $m$ and $k$. To evaluate these quantities, we introduce a family of Bell tests based on the Mermin-Svetlichny (MS) inequalities \cite{mermin,svetlichny}. Specifically, we compute the maximal value of the Mermin-Svetlichny (MS) expressions achieved by models where $n$ parties form $m$ groups and where $k$ parties broadcast their inputs. By comparing the amount by which quantum states violate the MS inequalities with our bounds, one thus obtains constraints on the values of $m$ and $k$ necessary to reproduce their nonlocal correlations. Since these criteria are based on Bell-like inequalities they can be tested experimentally.

A Bell-like test for a given number of groups could a priori depend on how the groups are formed, e.g., 2+2 in Fig.~1 $b)$ or 1+3 in Fig.~1 $c)$, and on which party belongs to which group. But, the tests that we present here depend only on the total number $m$ of groups, and not on how the parties are distributed within each group. Furthermore, in the measurement scenario that we consider in this work (restricted to ``correlation functions''), a communication model with $m$ disjoint groups is less powerful than a communication model with $k=n-m$ broadcasting parties. Yet, we find that the bounds on the MS expressions are identical in both cases.

As mentioned above, our results can be used to estimate the multipartite nonlocal content of quantum states. We carry out this analysis for GHZ-like and W states in the last part of this paper.

\paragraph{Definitions.}
We consider a Bell experiment involving $n$ parties which can each perform one out of two measurements. The outcomes of these measurements are written $a_j$ and $a'_j$ and can take the values $\pm 1$. Letting $M_1=a_1$, we define recursively the MS polynomials \cite{mermin,svetlichny,ww,collins} as
\ba
M_n&=&\frac12 (a_n+a_n')M_{n-1}+\frac12 (a_n-a_n')M_{n-1}'\label{defM}\\
\label{defM+}
M^{\pm}_n&=&\frac{1}{\sqrt{2}}\left(M_n\pm M'_n\right)\,,
\ea
where $M'_n$ is obtained from $M_n$ by exchanging all primed and non-primed $a_j$'s. $M^{+}_n$ and $M^-_n$ are equivalent under the exchange $\{a_j,a'_j\}\leftrightarrow \{-a'_j,a_j\}$ for any single party $j$, which corresponds to a relabeling of its inputs and outputs. The MS polynomials are symmetric under permutations of the parties.

We interpret these polynomials as sums of expectation values by identifying each term of the form $a_1\ldots a_n$ with the correlation coefficient $\langle a_1\ldots a_n\rangle$, which is the expectation value of the product of the outputs $a_1\ldots a_n$. The above polynomials can thus be interpreted as Bell inequalities. Their local bounds are known \cite{collins} to be $|M_n|\leq 1$ and $|M^\pm_n| \leq \sqrt{2}$, while the algebraic bounds (the maximal value achieved by an arbitrary nonlocal model) are easily found to be $|M_n| \leq 2^{\lfloor \frac{n}{2} \rfloor}$ and $|M^\pm_n| \leq 2^{\lfloor\frac{n-1}{2}\rfloor+\frac{1}{2}}$.

In the remainder of this paper, we shall be interested in the following family of polynomials:
%\begin{equation}\label{eq:Svetlichny}
$$S_n^m=
\begin{cases}
M_n & \text{for } \ n-m \text{ even}\\
M^+_n & \text{for } \ n-m \text{ odd}\,.
\end{cases}$$
%\end{equation}

\paragraph{Quantifying multipartite nonlocality through communication models.}
In a classical communication model, the $n$ parties have access to shared randomness and are allowed to communicate their inputs to {some subset of} the other parties. Given the information available to them, each party then produces a local output. We refer to \cite{jones} for a detailed  description of such models and a general formalism to analyze them. Here, as explained in the introduction, we define two families of models that depend on a parameter $m$ (or $k=n-m$) which quantify the extent of multipartite nonlocality.

$\bullet$ \emph{\textbf{Grouping}}: The $n$ parties are grouped into $m$ subsets. Within each group, the parties are free to collaborate with each other, but are not allowed to do so between distinct groups.

$\bullet$ \emph{\textbf{Broadcasting}}: Out of the $n$ parties, $k$ of them can broadcast their input to all other parties. The remaining $m=n-k$ parties cannot communicate their input to any other party.

In the framework of these two communication models, the values that can be reached by the MS polynomials are bounded as follows:

%
%Our objective now is to compute the maximal value of the MS expressions achieved by both models. These two models share the feature that there exist a special subset of $m$ parties such that none of the $n$ parties knows more than one input from this subset (this is obvious in the case of the broadcasting model; in the case of the grouping model, simply pick one party in each of the $m$ groups). As we will see shortly, this feature limits in a fundamental way the value of the $S_n^m$ expressions. This leads us to also consider the most general communication model with this property.
%
%This shows, as observed in \cite{jones} for the case $m=2$, that the structure of the MS inequalities allows to detect a stronger form of non-locality than the one induced by grouping. We will identify precisely the most general communication model associated with this stronger form of nonlocality.
%
%
%$\bullet$ \emph{Restrained-subset}: Among the $n$ parties, there is a subset of $m$ parties, such that none of the $n$ parties receive more than one input from this subset. The other parties are free to communicate as they wish. Note that the parties within the special subset of $m$ parties cannot receive inputs from any other party in the subset, as they already receive their own input.

\textbf{Theorem.} \emph{For the grouping and the broadcasting models,
\begin{equation}\label{bound}
|S_n^m| \leq 2^{(n-m)/2}\,.
\end{equation}
Moreover this bound is tight, i.e., for each model there exists a strategy that yields $|S_n^m|=2^{(n-m)/2}$ (in the case of the grouping model, this is true for any possible grouping of the $n$ parties into $m$ groups).
}

Before proving our theorem, let us elaborate on some comments. First of all, let us mention that for $m=2$, the results obtained in \cite{NL,collins} for the grouping model are recovered.

Note also that since we consider correlation functions only, the {grouping} model is weaker than the {broadcasting} model. Indeed, within each of the $m$ groups, we can assume that all parties send their input to one singled-out party, who can then produce the correct output for the entire group. This is because in correlation coefficients $\langle a_1\ldots a_n\rangle$, only the product of the outputs matters and not each output individually. We thus have singled out $m=n-k$ parties (one for each group), that do not need to send any input, but receive inputs from some of the other $k$ parties. This situation clearly involves less communication than the {broadcasting} model.

The fact that the same bounds hold for the two models is not trivial and is actually a special property of the MS expressions. Indeed, we have been able to construct inequalities that distinguish between these models.

%Note that the {restrained-subset} model is optimal for the MS expressions in the sense that any additional communication between the parties allows them to violate the bound \eqref{bound}. Indeed, if there is a party that receives two inputs from the restrained subset, this party knows the inputs of $n-m+2$ parties (the two from the restrained subset plus the ones from the $n-m$ unrestrained parties). It can thus reproduce the correct output for the corresponding group of $n-m+2$ parties. Assuming conservatively that the remaining $m-2$ parties produce each a local output that do not depend on the input of any other parties, the $n$ parties are thus separated into $m-1$ groups. Since the bound (\ref{bound}) is tight, the parties can achieve the bound for $m-1$ groups, which is strictly greater than the bound for a restrained subset of $m$ parties.

Note finally that since the above bounds are tight for both models, bounds for any intermediary model, in which for instance two parties would join to form a group and another one would broadcasts its input, can be readily computed; in this case it would correspond to $k=2$ or $m=n-2$.

\emph{A more technical remark}. The following remark and proofs will now be more technical. At first reading, one can simply skip this part and move right away to the discussion on the nonlocality of quantum states.

As observed in \cite{jones} for the case $m=2$, the structure of the MS inequalities allows to detect a stronger form of non-locality than the one induced by grouping. It is interesting to identify precisely the most general communication model associated with this stronger form of nonlocality.

%For that, we note that the two models above share the feature that there exists a special subset of $m$ parties such that none of the $n$ parties knows more than one input from this subset (this is obvious in the case of the broadcasting model; in the case of the grouping model, simply pick one party in each of the $m$ groups). Let us define the most general communication model with this property:

The common feature of the two above models that we exploit in our proof (see below) and that fundamentally limits the values of the $S_n^m$ expressions is that in both cases, there exists a special subset of $m$ parties such that none of the $n$ parties knows more than one input from this subset. This is obvious in the case of the broadcasting model; in the case of the grouping model, simply pick one party in each of the $m$ groups. Let us therefore define the most general (but less natural) communication model with this property:

$\bullet$ \emph{\textbf{Restrained-subset model}}: Among the $n$ parties, there is a subset of $m$ parties, such that none of the $n$ parties receive more than one input from this subset. The other parties are free to communicate as they wish. Note that the parties within the special subset of $m$ parties cannot receive inputs from any other party in the subset, as they already know their own input.

%This model also satisfies the bound (\ref{bound}). It is optimal for the MS expressions, in the sense that any additional communication between the parties allows them to violate (\ref{bound}).

This model also satisfies the bound (\ref{bound}); for the case $m=2$, the results of \cite{jones} are recovered. This model is optimal for the MS expressions, in the sense that any additional communication between the parties allows them to violate (\ref{bound}).

\emph{Proof of (\ref{bound})}. It is sufficient to prove (\ref{bound}) for our strongest model, i.e., for the {restrained-subset} model. Since the MS inequalities are symmetric under permutations of the parties, we can assume without loss of generality that the parties $1,\ldots,m$ are the ones in the restrained subset.

Consider first the case where $n-m$ is even, for which $S_n^m=M_n$. Applying twice the recursive definition (\ref{defM}), we get \begin{equation}
\begin{split}\label{mndec}
M_n =& \frac12(a_n a_{n-1}M_{n-2}'+a_n a_{n-1}'M_{n-2}\\
&+a_n' a_{n-1}M_{n-2}-a_n' a_{n-1}'M_{n-2}')\,.
\end{split}
\end{equation}
Using again twice (\ref{defM}) for $M_{n-2}^{(\prime)}$, we can replace $M^{(\prime)}_{n-2}$ as a function of $M_{n-4}^{(\prime)}$ in (\ref{mndec}). Iterating this process $\frac{n-m}{2}$ times, we end up with the following expression for $M_n$:
\begin{equation}
M_n=\frac{1}{2^{(n-m)/2}}\!\!\sum_{s_n,\ldots, s_{m+1}=0}^1 a_n^{s_n}\ldots a_{m+1}^{s_{m+1}}\,M_{m}^{s_n\ldots s_{m+1}}\label{Mbr},
\end{equation}
where $a_i^0=a_i$, $a_i^1=a_i'$, and where, depending on the value of $(s_n,\ldots,s_{m+1})$, $M_{m}^{s_n,\ldots,s_{m+1}}$ is equal to one of the polynomials $\{\pm M_{m},\pm M_{m}'\}$.

The MS polynomial  $M_{m}^{s_n\ldots s_{m+1}}$ is a function of the outputs $\{a_1,a_1'\ldots,a_m,a_m'\}$, i.e. $M_{m}^{s_n\ldots s_{m+1}}=M_{m}^{s_n\ldots s_{m+1}}(a_1,a_1'\ldots,a_m,a_m')$.
Among the parties $\{m+1,\ldots,n\}$ there exists a (possibly empty) subset $\{j_1,\ldots,j_l\}$ that do not receive any input from parties $2,\ldots,m$, but possibly from party 1. Define two effective outputs $A_1$ and $A_1'$ as $A_1 = a_1a^{s_{j_1}}_{j_1}\ldots a^{s_{j_l}}_{j_l}$ and $A'_1 = a'_1a^{s_{j_1}}_{j_1}\ldots a^{s_{j_l}}_{j_l}$. There also exist similar disjoint subsets for parties $2,\ldots,m$, for which we also define effective outputs $A_2,A'_2,\ldots,A_m,A'_m$. Then we can write
%By hypothesis, there exists among the parties $\{m+1,\ldots,n\}$, a subset $\{j_1,\ldots,j_l\}$ that do not receive any input from parties $2,\ldots,m$, but possibly from party 1. Define two effective outputs $A_1$ and $A_1'$ as $A_1 = a_1a^{s_{j_1}}_{j_1}\ldots a^{s_{j_l}}_{j_l}$ and $A'_1 = a'_1a^{s_{j_1}}_{j_1}\ldots a^{s_{j_l}}_{j_l}$. Introduce in a similar way effective outputs $A_2,A'_2,\ldots,A_m,A'_m$. Then we can write
\begin{equation*}\begin{split}\label{Meff}
a_n^{s_n}\ldots a_{m+1}^{s_{m+1}}\,&M_{m}^{s_n\ldots s_{m+1}}\\
&=M_{m}^{s_n\ldots s_{m+1}}(A_1,A_1'\ldots,A_m,A_m').
\end{split}\end{equation*}
Formally, $M_{m}^{s_n\ldots s_{m+1}}(A_1,A_1'\ldots,A_m,A_m')$ is a MS polynomial that involves $m$ parties isolated from each other, since the outputs $A_j,A'_j$ of party $j$ do not depend on the input of any of the other $m-1$ parties. It can therefore not exceed its local bound 1. Inserting this bound in (\ref{Mbr}), we find $|M_n|\leq 2^{(n-m)/2}$.

For odd values of $n-m$, we have to consider the polynomials $S_n^m=M^+_n$. Using the definitions (\ref{defM}) and (\ref{defM+}), one can show that $M^+_n$ has a similar decomposition as $M_n$ in (\ref{Mbr}). The same reasoning as before then leads to $|M^+_n| \leq 2^{(n-m)/2}$. \hfill$\Box$

\emph{Proof of the tightness of (\ref{bound})}. To prove that (\ref{bound}) is a tight bound, it is sufficient to prove that it can be reached by our weaker communication model, ie the {grouping} model (for any possible distribution of the $n$ parties into $m$ groups).

Let $G_i$ ($i=1,\ldots,m$) denote the $m$ groups into which the $n$ parties are split. For all group $G_i$ having an odd number $n_i$ of parties, there exists a strategy for the parties in $G_i$ to reach both algebraic bounds $|M_{G_i}|=|M'_{G_i}|=2^{(n_i-1)/2}$ at the same time. This is because the (tight) algebraic bound for $M_{G_i}^+$ is $2^{n_i/2}$ and Eq.~\eqref{defM+} tells us that in order to achieve it both $M_{G_i}$ and $M'_{G_i}$ must reach their algebraic limit. Similarly there exists a strategy for groups with even number of parties $n_i$ such that $|M_{G_i}^+|=|M_{G_i}^-|=2^{(n_i-1)/2}$. We shall thus associate $\{M,M'\}$ polynomials to odd groups and $\{M^+,M^-\}$ to even ones.

Consider two groups $G_i$, $G_j$, and their union $G_{ij}=G_i\cup G_j$. From the definitions \eqref{defM} and \eqref{defM+}, one can derive the decompositions:
%\begin{equation*}
%M_{G_{ij}}^{u}=\frac{1}{2}\left[M_{G_i}^{t_1}\left(M_{G_j}^{t_2}+M_{G_j}^{t_2'}\right)\pm %M_{G_i}^{t_1'}\left(M_{G_j}^{t_2}-M_{G_j}^{t_2'}\right)\right]
%\end{equation*}
\begin{align*}
M_{G_{ij}}&=\frac{1}{2}\left[M_{G_i}\left(M_{G_j}+M'_{G_j}\right)+M'_{G_i}\left(M_{G_j}-M'_{G_j}\right)\right]\nonumber\\
M_{G_{ij}}&=\frac{1}{2}\left[M^{+}_{G_i}\left(M^{+}_{G_j}+M^{-}_{G_j}\right)+M^{-}_{G_i}\left(M^+_{G_j}-M^-_{G_j}\right)\right] \label{decompM} \\
M^\pm_{G_{ij}}&=\frac{1}{2}\left[M^\pm_{G_i}\left(M_{G_j}+M'_{G_j}\right)\mp M^\mp_{G_i}\left(M_{G_j}-M'_{G_j}\right)\right]\nonumber\,.
\end{align*}
Similar relations are also obtained for $M'_{G_{ij}}$ since $(M_G^\pm)'=\pm M_G^\pm$.
Now inserting in the above relations the value attained by the strategies that we just mentioned for the two initial groups, one finds that their combined strategy can achieve $|M_{G_{ij}}|=|M'_{G_{ij}}|=2^{(n_i-1)/2}\,2^{(n_j-1)/2}$ or $|M^+_{G_{ij}}|=|M^-_{G_{ij}}|=2^{(n_i-1)/2}\,2^{(n_j-1)/2}$, depending on which set of polynomials are associated to the two initial groups.

Iterating this construction by joining groups successively 2 by 2, we find $|M_n|=\prod_{i=1}^m 2^{(n_i-1)/2}$ when there is an even number of even groups, and $|M^+_n|=\prod_{i=1}^m 2^{(n_i-1)/2}$ otherwise. Since the parity of the number of even groups is the same as the parity of $n-m$, there must exists a strategy which achieves $|S_n|=\prod_{i=1}^m 2^{(n_i-1)/2}=2^{(n-m)/2}$.\hfill$\Box$

\paragraph{Nonlocality of quantum states.}
Suppose that one observes a violation of the inequality $|S_n^m|\leq 2^{(n-m)/2}$. One can then conclude that in order to reproduce the corresponding nonlocal correlations in the framework of our communication models, the parties cannot be separated in more than $m-1$ groups, or that at least $k+1=n-m+1$ parties must broadcast their input. Thus, the above bounds on $S_n^m$ give us bounds on the multipartite character of the observed nonlocal correlations (an upper bound on $m$, or a lower bound on $k$).

\begin{figure}
\begin{center}
\includegraphics[width=0.36\textwidth]{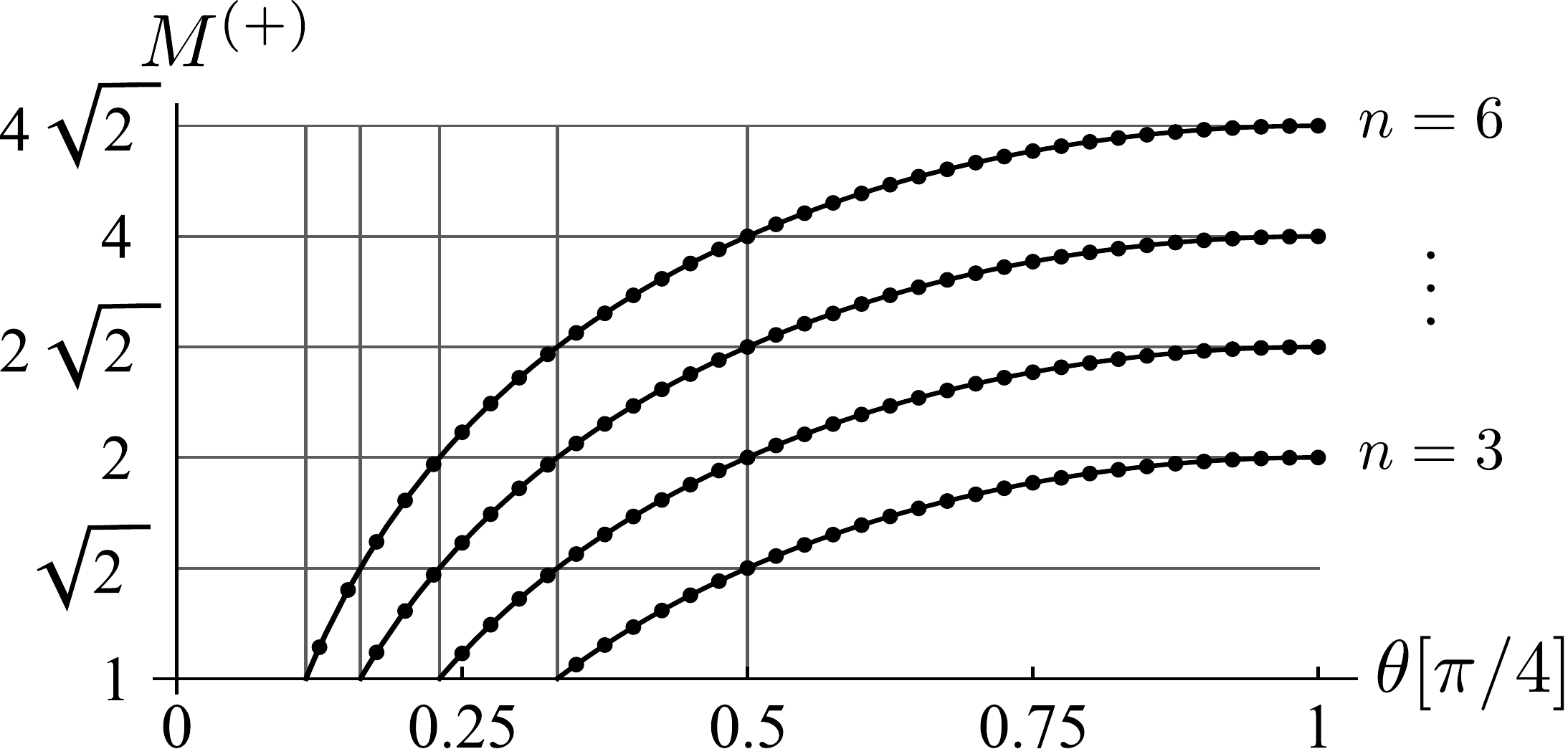}
\end{center}
\caption{Maximal values of $M_n$ and $M^+_n$ for partially entangled GHZ states for $3\leq n\leq 6$. The dots are values found by numerical optimization and the solid lines are the conjectured violation $M_n=M^{+}_n=2^{(n-1)/2}\sin 2\theta$ (valid only above 1 for $M_n$ and $\sqrt{2}$ for $M_n^+$).}
\label{fig:GHZM}
\end{figure}

Here we discuss the violation of the inequalities (\ref{bound}) for $n$-partite GHZ-like and W states. States in the GHZ family are defined as $\ket{GHZ_\theta}=\cos\theta\ket{00...0}+\sin\theta\ket{11...1} \label{def_GHZ}$. The maximal value of $M_n$ for these states was conjectured in \cite{scarani} to be $M_n=\max\{1,2^{(n-1)/2}\sin 2\theta\}$. Numerical optimizations (see Fig.~\ref{fig:GHZM}) induce us to conjecture that similarly $M_n^{+}=\max\{\sqrt{2},2^{(n-1)/2}\sin 2\theta\}$. Upon comparison with the bound (\ref{bound}), we conclude that all $n$-partite GHZ states with $\theta>\pi/8$ are maximally non-local according to our criterion (i.e., all parties must be grouped together or $n-1$ parties must broadcast their input to reproduce their correlations). Less entangled GHZ states, on the other hand, cannot be simulated if the parties are separated in more than $m-1$ groups or if fewer than $k+1=n-m+1$ parties broadcast their inputs whenever $\theta>\theta_c$ with $\sin 2\theta_c=2^{-(m-1)/2}$. Interestingly, $\theta_c$ is the same for all $n$.

\begin{figure}
\begin{center}
\includegraphics[width=0.34\textwidth]{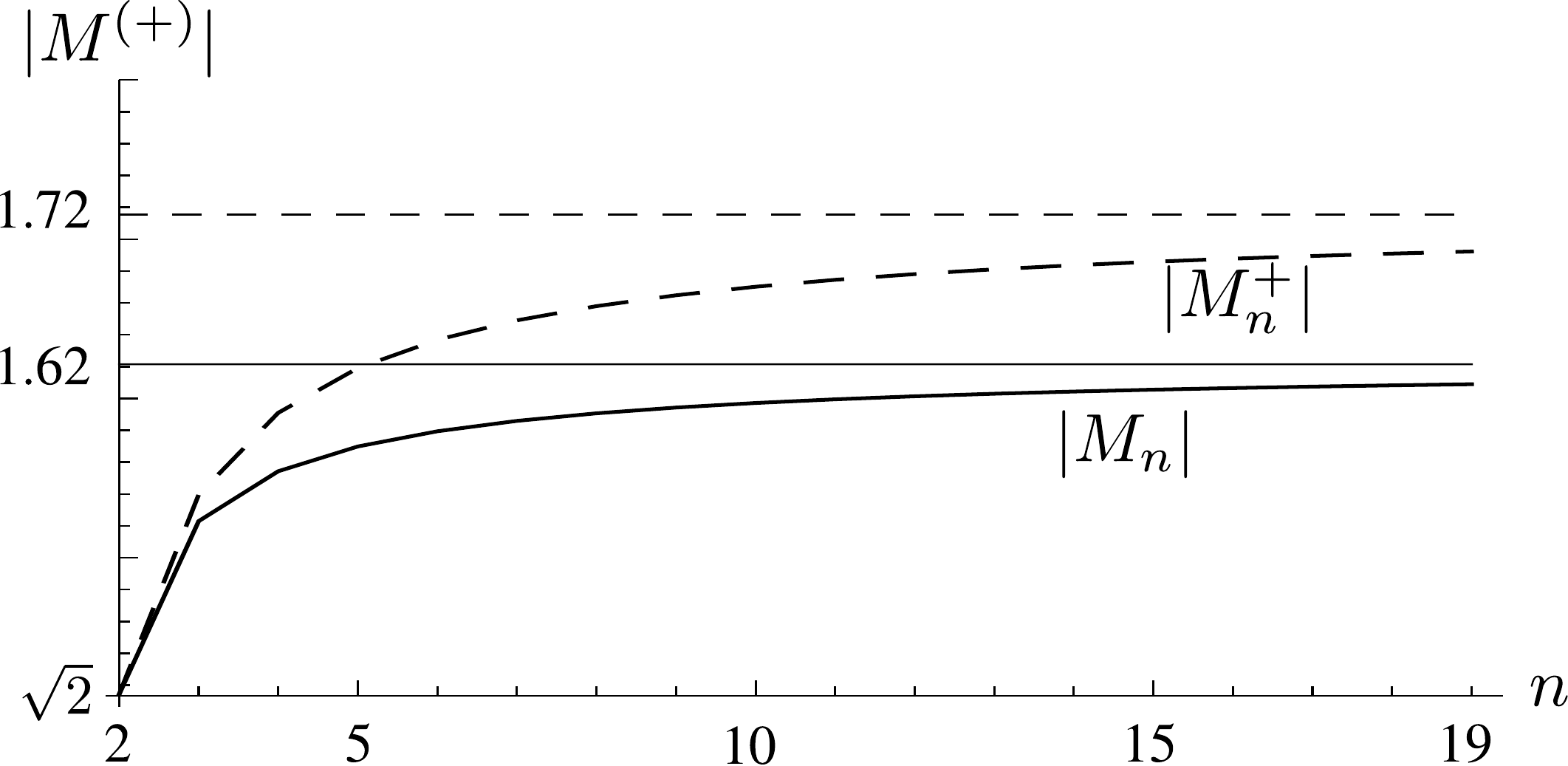}
\end{center}
\caption{Maximal values of $M_n$ (solid line) and $M^+_n$ (dashed line) for $n$-partite W states. The curves were obtained by a general numerical optimization for $n\leq 9$, and under the hypothesis that all parties use identical measurement settings for $10\leq n\leq 19$. The asymptotic values for $n\to\infty$ computed as explained in the text are also shown.}
\label{fig:WMM+}
\end{figure}

 Consider now the W states $\ket{W_n}= \frac{1}{\sqrt{n}} (\ket{10 \ldots 0}+\ldots +\ket{0 \ldots 01})\,$. Numerical optimizations suggest that the maximal values of the MS polynomials for these states are upper-bounded by a small constant for all $n$ (see Fig.~\ref{fig:WMM+}). To convince ourselves that this is indeed the case, we analyzed analytically the case where all pairs of measurement settings are the same for all parties. This is justified by the results of our numerical optimizations up to $n=9$, for which the optimal measurement settings can always be of this form. We thus introduce for all $n$ parties two measurement operators $A_0$ and $A_1$ represented by vectors $\vec a_i = (\sin \theta_i \cos \phi_i, \sin \theta_i \sin \phi_i, \cos \theta_i)$. One can show that as $n$ increases, the maximal value of $|M_n|$ or $|M_n^+|$ can be reached for $\phi_i=0$ and $\theta_i\to 0$. Assuming a power law for $\theta_i(n)$, one finds that it should be given by $\theta_i\sim c_i/\sqrt{n}$ at the maximum. After optimization of the constants $c_0$ and $c_1$ for both $M_n$ and $M_n^+$, we found that the asymptotic maximal values of the MS polynomials (under our assumptions, which we believe are not restrictive) are $|M_\infty| \simeq 1.62,\quad |M^+_\infty| =2\sqrt{2/e}$. Since $S_n^{n-1}>1$ for $n\geq3$, letting one party broadcast his input is not sufficient to reproduce the correlations of the W state.
However, we cannot reach the same conclusion for $k=n-m\geq 2$ broadcasting parties since the criterion (\ref{bound}) is not violated in this case. Similarly, for the {grouping} model it is not sufficient that two parties join together (so that the $n$ parties form $n-1$ groups) to simulate a W state, but we find no violation if they are separated in less than $n-1$ groups.

\paragraph{Conclusion.}
We have proposed in this paper two simple measures of multipartite nonlocality and have introduced a series of Bell tests to evaluate them. This represents a primary step towards a quantitative understanding of quantum nonlocality for an arbitrary number $n$ of parties.

While GHZ states exhibit a strong form of multipartite nonlocality according to our criterion, we have found that W states violate our inequalities only for small values of $k$. This suggest that $W$ states exhibit only a very weak form of multipartite non-locality. Or, it might actually be that other inequalities are necessary to quantify properly the nonlocality of W states. Finding which one of these possibilities is the correct one is an interesting problem for future research. Also, it would be interesting to analyze the non-locality of other kinds of multipartite quantum states with our criterions.

As suggested by the situation in entanglement theory, we do not expect our measures to be the only ways to quantify the multipartite content of nonlocality. It would thus be of interest to look for other ways to quantify multipartite nonlocality, based on other nonlocal models than the ones considered here.

Finally, let us stress that the criteria that we presented in this paper can be tested experimentally. It would thus be worth \mbox{(re-)considering} experiments on multipartite nonlocality in view of our results.

We acknowledge support by the Swiss NCCR \textit{Quantum Photonics} and the European ERC-AG \textit{QORE}.

\bibliography{references}

\end{document}